\documentclass[11pt]{article}

\usepackage{fullpage}
\usepackage{amsmath}
\usepackage{amsfonts}
\usepackage{multirow}
\usepackage{graphicx}
\usepackage{amsthm}

\newcommand{\ket}[1]{|{#1}\rangle}

\newcommand{\B}{{\cal B}}
\newcommand{\Sym}{{\rm Sym}}
\newcommand{\sign}{{\rm sign}}
\newcommand{\id}{\ensuremath{\mathbb I}}
\newcommand{\rn}{\ensuremath{\mathbb R}}
\newcommand{\rotspace}{\ensuremath{\rn^{3 \times 3}}}

\newcommand{\Cl}{\textsc{Clifford}}
\newcommand{\Cln}{\Cl\ensuremath{^\ast}}
\newcommand{\noise}{\ensuremath{\hat \theta}}
\newcommand{\noiseval}{\ensuremath{(6-2\sqrt{2})/7}}

\newcommand{\ip}[2]{\ensuremath{\langle #1, #2 \rangle}}

\newcommand{\C}{{\ensuremath{\cal C}}}
\newcommand{\R}{{\ensuremath{\cal R}}}

\newtheorem{mytheorem}{Theorem}
\newtheorem{mylemma}{Lemma}
\newtheorem{myclaim}{Claim}
\newtheorem{mycor}{Corollary}
\newtheorem{myob}{Observation}

\theoremstyle{remark}
\newtheorem{myrem}{Remark}

\begin{document}
\title{New Limits on Fault-Tolerant Quantum Computation}
\author{Harry Buhrman%
\thanks{Supported by the EU project RESQ IST-2001-37559 and 
the NWO vici project 2004-2009.}\\
{\small CWI and} \vspace*{-1.5mm}\\
{\small U of Amsterdam} \vspace*{-1.5mm}\\
{\small buhrman@cwi.nl}
\and
Richard Cleve%
\thanks{Supported in part by Canada's NSERC, CIAR, MITACS, and the US ARO.}\\
{\small U of Waterloo and} \vspace*{-1.5mm}\\
{\small Perimeter Institute} \vspace*{-1.5mm}\\
{\small cleve@cs.uwaterloo.ca} \\
\and
Monique Laurent
\thanks{Supported by the Netherlands Organization for Scientific Research NWO
639.032.203.}\\
{\small CWI, Amsterdam} \vspace*{-1.5mm}\\
{\small M.Laurent@cwi.nl}\\
\and
Noah Linden\thanks{Supported by the EU project RESQ IST-2001-37559 and UK EPSRC IRC in Quantum Information Processing.}
\\
{\small U of Bristol} \vspace*{-1.5mm}\\
{\small n.linden@bristol.ac.uk}\\
\and
Alexander Schrijver\\
{\small CWI and} \vspace*{-1.5mm}\\
{\small U of Amsterdam} \vspace*{-1.5mm}\\
{\small Lex.Schrijver@cwi.nl}\\
\and
Falk Unger$^*$\\
{\small CWI, Amsterdam} \vspace*{-1.5mm}\\
{\small unger@cwi.nl} }
\maketitle


\abstract{We show that quantum circuits cannot be made
fault-tolerant against a depolarizing noise level of
$\noise = (6-2\sqrt{2})/7 \approx 45\%$, thereby improving on
a previous bound of $50\%$ (due to Razborov \cite{R03}).
Our precise quantum circuit model enables {\em perfect} gates from
the Clifford group (CNOT, Hadamard, $S$, $X$, $Y$, $Z$) and
arbitrary additional one-qubit gates that are subject to depolarizing
noise $\noise$.
We prove that this set of gates cannot be universal for arbitrary
(even classical) computation, from which the upper bound on
the noise threshold for fault-tolerant quantum computation follows.}


\section{Introduction}
In the past decade, quantum computing has attracted much attention
because of its ability to efficiently solve problems for which
no efficient classical algorithms are known.
Significant research efforts are dedicated to physically realizing
quantum computers.
A fundamental problem is to cope with noise, which creates major
difficulties in storing and operating on quantum states reliably.
A key advance was the realization that quantum error correcting
codes~\cite{Shor95,Steane96a} exist and fault-tolerant quantum
computation~\cite{Shor96} is possible for a number of 
reasonable error models.
Subsequent results have 
improved on the first
fault-tolerant schemes, proving better and better bounds on the
noise tolerable in quantum computation (e.g.~\cite{CS96,AharonovB97}).
Recent results suggest that fault-tolerant quantum computation is
possible with gates that have as much as $3\%$ of depolarizing
errors \cite{Knill04}, but there is no rigorous proof so far.

In this paper we will concentrate on the opposite task of proving
that, for certain noise levels, quantum computation is
{\em impossible}.
Our main result is as follows:
Let $\Cl$ be the set of all (noiseless) Clifford gates
\begin{equation}
\label{Cliffords}
\mbox{CNOT$^1_2$}
=\left(\begin{array}{cccc}
1 & 0 & 0 & 0 \\
0 & 1 & 0 & 0 \\
0 & 0 & 0 & 1 \\
0 & 0 & 1 & 0
\end{array}
\right)~~~~
H=\frac{1}{\sqrt{2}}\left(\begin{array}{cr}
 1 & 1  \\
 1 & -1
\end{array}
\right)~~~~
S=\left(\begin{array}{cc}
1 & 0  \\0& i
\end{array}
\right)
\end{equation}
The Gottesman-Knill Theorem says that this set of gates can be efficiently 
simulated classically (see also \cite{AaronsonGottesman04}), so they are 
probably not universal for quantum computation. On the other hand, 
it is known that $\Cl$ together with {\em any} other
one-qubit gate, not generated by the gates in $\Cl$, form a
universal set of gates for quantum
computation~\cite{Solovay00,NebeRainsSlaon01}.
We show however, that
such additional one-qubit gates should not be too noisy.
More precisely, let $\Cln$ be  $\Cl$ augmented with arbitrary one-qubit
gates with depolarizing error at least $\noise=\noiseval\approx 45\%$.
Then this set of gates is no longer capable of computing arbitrary
functions and thus is not universal.
In other words, fault-tolerant quantum
computation cannot be performed if there is this level of noise.
Additionally we show that, among 
all one-qubit gates that augment $\Cl$, the so-called $\pi/8$-gate 
(see end of Section~\ref{sec:main}) is the type of gate that requires 
the most noise to render it incapable of universal quantum computation 
by our approach. That is, for other augmenting gates (e.g., $\pi/16$-gates), 
our approach will yield stronger bounds on the tolerable level.
Our results also yield a simple proof that not all classical functions 
can be computed using Clifford gates (complementing results in 
\cite{AaronsonGottesman04}).
In particular, in Corollary \ref{corParity}, we show 
that a boolean function which can be computed by Clifford circuits 
can be written as the parity of a subset of input bits.

The main idea of our approach is as follows. Assume we have a Clifford
circuit $C$ with $n$ classical input bits $x=x_1,\ldots,x_n$ and one
dedicated output qubit that, when measured in the computational basis,
yields the output of the computation of $C$ on $x$. Suppose now that
the input is partitioned over two parties,
Alice and Bob, such that Alice has $k$ bits
of $x$ and Bob has $n-k$ bits.  We first show how Alice, with the help
of Bob, can compute the value of $C$ on $x$ with just a single
classical bit of communication (Lemma \ref{lemmaMain}). From this it
follows that Clifford circuits can at the very best compute only those
functions that require for any partition of the inputs a single bit of
communication, and it is well known that many functions require more
than one bit of communication. Next, we show in Lemma \ref{lemmaSimulate} how
probabilistic mixtures of Clifford gates can be used to simulate any
single qubit unitary gate, that has noise \noise ($\approx 45\%$). 
 The proof of our
Lemma relies on solving an optimization problem related to the
Clifford polytope, defined as the convex hull of the set $\C\subseteq
\rotspace$ of Clifford rotation 
matrices in $\rn^3$. Here, the matrices $\C$
are essentially the one-qubit Clifford gates in Bloch sphere
representation.

Combining Lemmas~\ref{lemmaMain}
and~\ref{lemmaSimulate}, we get that all circuits with \Cln-gates and with
respect to any distribution of the inputs can
be computed by Alice and Bob  with a single  bit of communication (Lemma
\ref{lemmaqMain}).  Using the fact that there are functions which require
communication  more than one bit, we get our main result
(Theorem \ref{theMain}): The set of gates in $\Cln$ cannot be universal.
We also generalize our result to the case that the inputs
are quantum states.

The
idea that a noisy 1-qubit gate can be simulated by a probabilistic
mixture of Clifford appeared first in Virmani 
\textit{et al.}~\cite{VirmaniHuelgaPlenio04}. The approach we take
here  though is an extension to quantum fault tolerant computation of the work
by Brassard et al~\cite{BrassardBuhrmanLindenMethotTappUnger05}, where
they exhibit  an upper bound on the noise threshold for classical
fault tolerant computation, using
lower bounds on quantum communication complexity and the non-local
CHSH correlation.  

We want to point out that section \ref{secMain} can be read
independently of the preceding section. It shows that gates from \Cln,
together with all stabilizer operations and classical co-processing are
classically simulatable and thus probably not quantum-universal.



\subsection{Related Work}
There are only a few other results concerning the limits of fault-tolerant 
computation.
These are not all strictly comparable to each other and our result; 
nevertheless, we review them and make some comparisons.
See the introduction of \cite{R03} for some remarks that motivate the 
analysis of thresholds for fault-tolerant quantum computation.

The first results on upper bounds of the
threshold decoherence rate were obtained by showing that quantum
computers with faulty gates can be simulated efficiently on a
classical computer. The first to prove one of these results were
Aharonov and Ben-Or \cite{AB96}, with the value $97\%$ for the noise.
Later Harrow and Nielsen \cite{HN03} showed that if $74\%$ of
depolarizing noise is applied to each output qubit of each gate, 
then (faulty) two-qubit gates cannot produce entanglement. 
They concluded that circuits containing only one- and two-qubit gates 
with depolarizing noise at least $74\%$ can be simulated efficiently 
on a classical computer. 

An improvement of this is due to Virmani \textit{et al.}~\cite{VirmaniHuelgaPlenio04} who show that the set consisting of CNOT with depolarizing noise at least $67\%$ and arbitrary 1-qubit gates is efficiently simulable classically. In this paper they also introduce the interesting idea that sufficiently noisy 1-qubit gates can be simulated by Clifford gates; we build on and extend this idea in this paper. We note however, that their strongest results are for a restricted class of gates (ones which are diagonal in the computational basis) and dephasing or worst-case noise. They prove that $(\sqrt{2}-1)/\sqrt{2}\approx 29 \%$ dephasing noise is enough to make these diagonal gates a mixture of Clifford operations\footnote{They define dephasing noise as $\rho \mapsto 1/2(\rho + Z \rho Z)$.}.  We extend their results by considering all 1-qubit gates. Note also that dephasing noise is only symmetric around the $z$-axis, which is natural when considering diagonal gates. Our noise bounds are with respect to depolarizing noise, which is symmetric in all directions, and hence appropriate when considering arbitrary one-qubit unitaries.

Note that all these results do not exclude the
possibility that quantum circuits with high noise can still do 
universal classical computations; our results imply this.

The only prior result of this latter type is due to Razborov \cite{R03}, 
where a $50\%$ upper bound on the noise threshold for depolarizing noise 
on qubits for circuits with two-qubit gates is obtained (and a weaker 
bound for $k$-qubit gates).
The argument in \cite{R03} is essentially that, at this noise level, any 
superlogarithmic-depth quantum circuit (with constant error rate per 
qubit per time step) will be overwhelmed by the noise and produce a 
statistically meaningless outcome.
Thus, under the complexity theoretic assumption $BQP \neq QNC^1$, there are sets in $BQP$ which can be computed with this noise level. 
We note that it is shown in \cite{CleveW00} that in fact log-depth quantum 
circuits can perform interesting feats, including efficient integer 
factorization (if combined by classical polynomial-time pre- and 
post-processing).
Our error model is in most respects weaker than that of \cite{R03} (since 
our qubit errors are only occurring at the completion of non-\Cl\ gates) 
and our bound of $\approx 45\%$ is below 50\%.
In fairness, there is a sense in which the bound in \cite{R03} is stronger: 
it permits arbitrary (noisy) two-qubit gates; whereas, our only two-qubit 
gates are (perfect) CNOT gates.

Finally, we note that our work is related to, and partly stimulated by, 
the circle of ideas surrounding measurement-based quantum computation 
that was largely initiated by \cite{GottesmanC99,RaussendorfB01}.


\section{Preliminaries and Notation}
\label{secNot}
 $E_{ij}$ is the all-zero matrix, except for the entry $i,j$ which is equal to $1$. We also write $+$ for $+1$ and $-$ for $-1$.  For matrices $A,B \in \rotspace$ we define the inner product $\ip{A}{B}$ as:
$$\ip{A}{B}= tr(A^TB)=\sum_{i,j \in \{1,2,3\}} a_{ij}b_{ij}.$$ The following fact is used repeatedly: $\ip{A}{BC}=\ip{B^TA}{C}$ for $A,B,C \in \rn^{3 \times 3}$.

By $A^\ast$ we denote the conjugate transpose of matrix $A$.

An \emph{$n$-qubit state} (or density matrix) $\rho$ is a matrix $\rho \in \mathbb{C}^{2^n \times 2^n}$ with the properties $tr(\rho)=1$, $\rho=\rho^*$ (Hermiticity) and $\rho$ is positive semi-definite. An \emph{$n$-qubit operation} (or \emph{gate}) is a unitary matrix $U \in \mathbb{C}^{2^n \times 2^n}$, i.e., $U^*U=\id$. For such $n$-qubit state $\rho$ and $n$-qubit operation $U$ the application of $U$ to $\rho$ results in the state $U\rho U^*$.


\subsection{Bloch-vector representation}
\label{subsecBloch}
In our further analysis it will be convenient to use the Bloch-sphere representation of 1-qubit states and 1-qubit operations, which we review now (see e.g. Section 4.2 and Chapter 8 in \cite{NC00}).

For $\mathbf{r} \in \rn^3$ define $\mathbf{r} \cdot \mathbf\sigma=r_x X+ r_y Y + r_z Z$, where $\mathbf \sigma=(X,Y,Z)$ is the vector of Pauli matrices
\begin{equation}
\label{Paulis}
X=\left(\begin{array}{cc}
0 & 1  \\1& 0
\end{array}
\right)~~~
Y=\left(\begin{array}{cc}
0 & -i  \\i& 0
\end{array}
\right) ~~~
Z=\left(\begin{array}{cc}
1 & 0  \\0& -1
\end{array}
\right).
\end{equation}
Then, all 1-qubit density matrices $\rho$ can be uniquely written in the form
$$\rho = \frac{\id + \mathbf{r} \cdot \mathbf\sigma}{2}=\frac{\id + r_x X+ r_y Y + r_z Z}{2},$$
where $\mathbf r \in \rn^3$ and $||\mathbf r||= \sqrt{r_x^2+r_y^2+r_z^2}\leq 1$. We call $\mathbf r$ the \emph{Bloch vector} of $\rho$.

For $\mathbf{n} \in \rn^3$ with $ ||\mathbf n||=1$ and $\theta \in \rn$ we define
$$U_{\mathbf n}(\theta)=\exp(-i\theta\mathbf n\cdot \mathbf\sigma/2)=\cos(\theta/2)\id -i \sin(\theta/2) \mathbf n \cdot \mathbf\sigma.$$ We first note that $U_{\mathbf n}(\theta)U_{\mathbf n}(\theta)^\ast=\id$, i.e., $U_{\mathbf n}(\theta)$ is unitary. Second, let the result of the quantum operation $U_{\mathbf n}(\theta)$ applied to state  $\rho=\id/2 + \mathbf{r} \cdot \mathbf\sigma/2$ be $ \rho'=U_{\mathbf n}(\theta)^\ast \rho U_{\mathbf n}(\theta)=\id/2 + \mathbf{r'}\cdot \mathbf\sigma/2 $. Then $\mathbf r'$ is the image of rotating $\mathbf r$ around $\mathbf n$ by an angle $\theta$. Third, all 1-qubit unitaries $U$ can be written as
$$U= U_{\mathbf n}(\theta)$$
with $\mathbf n \in\rn^3, \theta \in \rn$ and $||\mathbf n||=1$ (ignoring an unimportant phase factor $\alpha \in \mathbf{C}$ with $|\alpha|=1$).

Thus, one-qubit states and unitaries are isomorphic to vectors ,resp., rotations in $\rn^3$.  The set of all rotations in $\rn^3$ is the group $SO(3)$.\footnote{This group will play a prominent role in the proof of Lemma \ref{lemmaSimulate}, where some more notation can be found.} We introduce some notation reflecting this isomorphism.
For unitary $U \in \mathbb{C}^{2 \times 2}$ we let $R_U \in SO(3)$ be the corresponding rotation matrix. We get a reverse operation (up to phase factors) by fixing one mapping $f: SO(3) \rightarrow \mathbb{C}^{2 \times 2}$ with the property that for all unitary $U \in \mathbb{C}^{2 \times 2}$ it holds $f(R_U)=\alpha U$ for some $\alpha \in \mathbb{C}$, $|\alpha|=1$. We then write $U_R=f(R)$.

This can be extended to probabilistic mixtures of quantum operations. Let $\{p_i\}$ be a probability distribution, i.e., $\sum_i p_i=1$ and $0 \le p_i$, and let $U_i \in \mathbb{C}^{2 \times 2}$ be a 1-qubit unitary with corresponding Bloch representation $R_i \in \rotspace$. Then the quantum operation $E$ in which each $U_i$ is applied with probability $p_i$ has Bloch-representation $R_E=\sum_ip_iR_i$.


\subsection{Noise}
\label{subsecNoise}
There are several models of noise considered in the literature. The most common one, which we consider too, is \emph{depolarizing noise}.
A 1-qubit state $\rho$ to which depolarizing noise $p$ is applied, becomes
$$(1-p) \rho + p \id/2.$$
Thus, with probability $1-p$ the state is not changed and with probability $p$ the state is replaced with the completely mixed state.

It is not hard to see that applying depolarizing noise $p$ to $\rho = \id/2 + \mathbf{r} \cdot \mathbf\sigma/2$ yields  $\rho' = \id/2 + \mathbf{r'} \cdot \mathbf\sigma/2$, with $\mathbf{r'}=(1-p)\mathbf{r}$. So, this noise shrinks the Bloch vector of a state to $(1-p)$ of its original length.

We say that a 1-qubit gate implements the unitary $U$ with noise $p$ if it transforms states $\rho$ into
\begin{equation}
\label{unitaryNoise}
(1-p) U \rho U^\ast +p \id/2.
\end{equation}
This {quantum operation} can be seen as a two-stage process, in which first $U$ and then depolarizing noise is applied. Let $R_U \in \rotspace$ be the rotation matrix corresponding to the unitary $U$. Then this noisy quantum operation has Bloch-representation $(1-p)R_U$, i.e., it rotates a Bloch vector and scales it by a factor $1-p$.

For 1-qubit gates and depolarizing noise, the two representations are (up to unimportant global phase factors) equivalent.
(See Section 8.3 in \cite{NC00} for more details.)


\subsection{Clifford group}
\label{subsecClifford}
The ($n$-qubit) Clifford group contains all unitary operations that can be written as a product of tensor products of $S,H$ and $CNOT$ (see Eq. (\ref{Cliffords})).
The Clifford group contains also all Pauli operators $X,Y,Z$. We let \Cl\ be the set of all Clifford gates. Let \Cln\ be the set of gates consisting of \Cl\ and arbitrary 1-qubit gates which have depolarizing noise at least $\noise=\noiseval$.

For a state with Bloch vector $\mathbf r$ we get:
$$S\left(\frac{1}{2}\id  +\frac{r_x}{2}X+\frac{r_y}{2}Y+\frac{r_z}{2}Z \right)S^*= \frac{1}{2}\id  -\frac{r_y}{2}X+\frac{r_x}{2}Y+\frac{r_z}{2}Z $$
Let $R_S$ be the Bloch representation of $S$. Then $R_S$ rotates Bloch vectors  around the $z$-axis by $\pi/2$. In particular, the $x$-axis is mapped to $-y$ and $y$ to $x$. For the Hadamard-gates we similarly have
$$H\left(\frac{1}{2}\id  +\frac{r_x}{2}X+\frac{r_y}{2}Y+\frac{r_z}{2}Z \right)H^*= \frac{1}{2}\id  +\frac{r_z}{2}X-\frac{r_y}{2}Y+\frac{r_x}{2}Z .$$
So the Bloch representation $R_H$ of $H$ negates
 the $y$-coordinate of a Bloch vector and swaps the $x$ and $z$-coordinates, i.e., it is a rotation by $\pi$ around the axis $1/\sqrt{2}(1,0,1)$.

We define $\C$ as the set of matrices which can be generated from $R_S$ and $R_H$. A $C \in \C$ is called a \emph{Clifford (rotation) matrix}. It is not hard to see that $\C$ contains exactly those rotations which map axes to axes (or their opposite). Those $C$ have in each row and column exactly one non-zero entry, which must be either $+1$ or $-1$, and $det(C)=1$. Note that $\cal C$, being isomorphic to the 1-qubit Clifford group, is a group under matrix multiplication.
Examples of Clifford matrices are
$$\left(
\begin{array}{ccc}
1 &0 &0\\
0 &1 &0\\
0 &0 &1
\end{array}
\right),
\left(\begin{array}{ccc}
1 &0& 0\\
0 &-1& 0\\
0 &0 &-1
\end{array}
\right),
\left(\begin{array}{ccc}
1 &0& 0\\
0 &0& 1\\
0 &-1 &0
\end{array}
\right)
$$

In Appendix \ref{secFacets} we need (and explain) more details about Clifford rotation matrices.


\subsection{Communication Complexity}
The setting for this is the following: Assume two separated parties, Alice and Bob, where Alice is given
$x \in \{0,1\}^{m_A}$ and Bob $y \in \{0,1\}^{m_b}$, want to compute $f(x,y)$ for some fixed function
$f: \{0,1\}^{m_A}\times \{0,1\}^{m_B} \rightarrow \{0,1\}$.
We want that at least one party learns the result $f(x,y)$.
In order to achieve this they can communicate  bits, according to a predefined protocol.
The \emph{deterministic communication complexity}  $C_f(n)$ of $f$ is the smallest number
$c$ such that each protocol which always computes the correct result, needs at least $c$
bits of communication for at least one input $x,y \in \{0,1\}^n$.

It is well-known that there are functions $f$ where $C_f(n)$ is $n$, for example the inner product
function (see \cite{KN:comm}).
The above can be extended to \emph{randomized} communication, where the parties are additionally
provided with a source which sends a sequence of random bits to Alice and the same sequence to Bob.
The final result only has to be correct with some probability $1- \epsilon$ for $\epsilon<1/2$. The minimum number of bits needed to be communicated such that the output is correct with probability at least $1- \epsilon$ is denoted by $C^\epsilon_f(n)$.
However, also in this randomized setting there are ``hard'' functions.
For example, it is known that the inner product function has randomized communication
complexity $n-O(\log(1/\delta))$, if the outputs have to be correct with probability at
least $1/2-\delta$ (see also \cite{KN:comm}).

We will also slightly abuse notation and define communication complexity for functions $f : \{0,1\}^n \rightarrow \{0,1\}$ which depend only on one input string. For any $S \subseteq \{1,\dots,n\}$ let $C_f(n,S)$ be the communication complexity of $f$ if the bits with indices in $S$ are given to Alice and all others to Bob. We then set $C_f(n)=\max_{S \subseteq \{1,\dots,n\}}C_f(n,S)$.


\section{The power of Clifford circuits}\label{sec:main}
\label{secMain}
We are now ready to prove the main lemma, which explains the idea of simulating Clifford circuits.
\begin{mylemma}
\label{lemmaMain}
Let $f : \{0,1\}^n \rightarrow \{0,1\}$ be a function that is
computable with unbounded error\footnote{That means, that the output is only correct with probability greater than $1/2$, but can go arbitrarily close to $1/2$ as $n$ grows.} by a quantum circuit $C$ that uses only gates from \Cl\ , ancillas initialized to $|0\rangle$ and one single-qubit measurement in the computational basis, which determines the output. Then the deterministic communication complexity $C_f(n)$ is at most one bit.
\end{mylemma}
\begin{proof}
We begin by noting that each qubit can be represented by two \textit{shares}: a 
\textit{classical share} consisting of two bits, and a \textit{quantum share} 
consisting of one qubit .
When the classical share is $ab$ and the quantum share $|\psi\rangle$, then the logical qubit that the shares encode is $X^a Z^b|\psi\rangle$.

Assuming that a set of qubits is encoded in this manner, the operations $H$, $S$,
and CNOT can be applied to the logical qubits by separately performing operations
on the shares that encode them (i.e., the logical qubits do not have to be reconstructed). The reason why this works is because for any Clifford operation $C=H, S, \mbox{CNOT}^1_2$ and any tensor product of Pauli operators $P_1$ there is a tensor product of Pauli operators $P_2$ with $CP_1=P_2C$.
For example, to apply $H$ to a logical qubit, the two bits that make up its classical
share are swapped and $H$ is applied to its quantum share.
This works correctly because
\begin{eqnarray}
H X^a Z^b |\psi\rangle& =&  H X^a H H Z^b H H |\psi\rangle\\
& = & Z^a X^b H |\psi\rangle\nonumber \\
& = & (-1)^{a \wedge b} X^b Z^a H|\psi\rangle, \nonumber
\end{eqnarray}
and $(-1)^{a \wedge b}$ is an irrelevant global phase.

To apply $S$ to a logical qubit, the $b$-part of the classical share is updated to $b:=a \oplus b$ and $S$ is applied to its quantum share.
This case can be verified by noting that
\begin{eqnarray}
S X^a Z^b |\psi\rangle& =& i^a X^a S Z^a Z^b |\psi\rangle\\
& = & i^a X^a Z^{a\oplus b} S|\psi\rangle, \nonumber
\end{eqnarray}
where we note that $i^a$ is a global phase.

To simulate the application of $\mbox{CNOT}^1_2$ gate\footnote{control } on two logical qubits, with
classical shares $a_1 b_1$ and $a_2 b_2$, we update $a_2:=a_1 \oplus a_2$, $b_1:=b_1 \oplus b_2$ and
$\mbox{CNOT}^1_2$ is applied to the two quantum shares.
In this case, we omit the details but note that the correctness can be verified
using the identities
\begin{eqnarray}
\mbox{CNOT}^1_2 (X \otimes I) & = & (X \otimes X) \mbox{CNOT}^1_2 \\
\mbox{CNOT}^1_2 (I \otimes X) & = & (I \otimes X) \mbox{CNOT}^1_2 \nonumber \\
\mbox{CNOT}^1_2 (Z \otimes I) & = & (Z \otimes I) \mbox{CNOT}^1_2 \nonumber \\
\mbox{CNOT}^1_2 (I \otimes Z) & = & (Z \otimes Z) \mbox{CNOT}^1_2. \nonumber
\end{eqnarray}

We first describe a \textit{probabilistic} communication protocol for $f$.
Alice operates on the classical shares while Bob operates on the quantum shares.
The initial shares are easy to construct: for each of Alice's input qubits $\ket{x_j}$,
Alice sets her classical share to $a_j:=x_j$,$b_j:=0$ and Bob sets his quantum share to $\ket{0}$;
for each of Bob's input bits $y_j$, Alice sets her classical share to $a_j=b_j:=0$
and Bob sets his quantum share to $\ket{y_j}$.
In this manner, Alice and Bob can simulate the execution of circuit $C$ on input
$\ket{x}\ket{y}\ket{0\dots 0}$ \textit{without any communication} to obtain the
shares of the output qubits of $C$.
For Bob to obtain the measured output qubit, Alice sends the first bit of
her classical share, $a_1$, to Bob, who applies $X^{a_1}$ to his quantum share
and measures it (Alice need not send $b_1$, the second bit of the classical
share, since Bob is performing a measurement in the computational basis).

Finally, to obtain a \textit{deterministic} communication protocol for $f$, we note
that Bob need not actually manipulate quantum information; rather, he can simulate
his quantum registers and his operations with high enough precision on a classical computer.
Then, upon receipt of the classical bit from Alice, he can exactly determine
the output probabilities of his measurement to determine which outcome is more likely.
\end{proof}

 From Lemma \ref{lemmaMain} we get that the set of functions computable with Cliffords gates is very limited and far from being universal.
\begin{mycor}
\label{corParity}
Every function $f :\{0,1\}^n \rightarrow \{0,1\}$ which can be computed by a Clifford circuit, can be written in the form
$$f(x_1\dots x_n)=c \oplus \bigoplus_{j \in S}x_j,$$
where $S \subseteq [n]$ is a subset of the input bits not depending on the input bits and $c \in \{0,1\}$. 
\end{mycor}
\begin{proof}
Let $f :\{0,1\}^n \rightarrow \{0,1\}$ be a function which can be computed by a Clifford circuit $C$. Then we can simulate this circuit as in Lemma \ref{lemmaMain}, where we give Alice the whole input, i.e.,  $m_A=n$ and $m_B=0$. 

Inspecting the proof of Lemma \ref{lemmaMain} we see that in each step Alice always updates her $a_i$'s and $b_i$'s by computing the parity of two bits. So, the final bit she sends over, say $a_i$, is just the parity of some of the input bits. Thus we can write $a_i=\bigoplus_{j \in S}x_j$, for some $S \subseteq [n]$.
Bob initializes all his quantum bits to $|0\rangle$, so he starts with the state $|\psi^0\rangle=|0\dots 0\rangle$. Further, Bob just applies the circuit $C$ to his state and measures the $i$-th qubit of $X^{a_i}C|\psi^0\rangle$ in the computational basis.

It is known that the probability for measuring $1$ in a Clifford circuits is either $0$, $1/2$ or $1$ (see \cite{NC00} page 463). It cannot be $1/2$ in our case, because that would mean that the circuit does not compute $f$. So, measuring the $i$-th bit of  $C|\psi^0\rangle$ yields a bit $c \in \{0,1\}$ with certainty. But this means that $f(x)=c \oplus a_i= c \oplus  \bigoplus_{j \in S}x_j$.
\end{proof}
We mention that Aaronson and Gottesman proved \cite{AaronsonGottesman04} that there is a log-space machine which transforms a Clifford circuit $C$ into a classical circuit $C'$ consisting only of CNOT and NOT gates, with the property that $C$ accepts the all zero state $|0\rangle^{\otimes n}$ iff $C'$ accepts the (classical) all zero input. Our corollary extends this slightly: For every Clifford circuit $C$ computing a boolean function, there is an equivalent (for classical inputs) classical circuit which uses only NOT- and CNOT-gates.
Using the result from \cite{AaronsonGottesman04} we see that we can compute the bit $c$ in the proof of Corollary \ref{corParity} in log-space and it is also clear that the circuit Alice uses to compute $a_i$ can be computed in log-space.


\paragraph{Quantum inputs}
\label{remQuantum}
Lemma \ref{lemmaMain} can actually be extended to she case where Alice and Bob get quantum states as 
inputs and they are provided with entanglement. It is no problem for Bob to start 
with a quantum state as an input. For Alice we do the following. We let her teleport 
her quantum input to Bob bit by bit, using the standard scheme for teleportation 
(see e.g. \cite{NC00}). When Alice teleports a qubit, which corresponds to the 
$i$-th input qubit of the circuit $C$ to be simulated, she measures two classical bits. 
Now, if she does not send these to Bob, but rather initializes her $a_i,b_i$ with 
these bits, Alice and Bob obtain the correct representation for qubits of $C$ as 
in Lemma \ref{lemmaMain}. Since the inner product function has communication complexity 
$\Omega(n)$ even in the presence of entanglement \cite{CDNT:InnerProduct} we see that 
Theorem \ref{theMain} is also true for quantum inputs.

Note however, that now Bob can no longer compute the result of the circuit with 
certainty, because he does not know his input state. The correctness probability 
will only be the same as that of the simulated circuit $C$ itself.

\begin{myrem}
\label{remMoreoutput}
It is straightforward to extend these results to functions with $m$ output bits, if the communication complexity of the function is also higher than $m$, resulting in a scheme that uses $m$ bits of communication.
\end{myrem}

\section{Simulating unitaries}
\label{secSim}

We want to extend Lemma \ref{lemmaMain}, by replacing \Cl\  with {\Cln}. To do that we first show how one can simulate arbitrary 1-qubit gates with depolarizing noise $\noise=\noiseval$ with a probabilistic mixture of Clifford operations.
\begin{mylemma}
\label{lemmaSimulate}
Let $U$ be a 1-qubit unitary and $E_U$ be the following noisy version of it
$$\rho \mapsto E_U(\rho)=(1-\noise) U\rho U^* + \noise \id/2,$$
for any $\rho \in \mathbb{C}^{2 \times 2}$. Then there is a probability distribution $\{p_C\}$ over $\C$ such that for all $\rho \in \mathbb{C}^{2 \times 2}$ we have
$$E_U(\rho)=\sum_{C \in \C}p_C U_C \rho U_C^*$$
and $U_C$ is a Clifford operation corresponding to the Clifford rotation matrix $C$.
\end{mylemma}
\begin{proof}
Using Section \ref{subsecBloch} the lemma can be reformulated equivalently in Bloch representation:
For any $S \in SO(3)$ there is a probability distribution $\{p_C\}$ over $\C$  such that
\begin{equation}
\label{repr}
(1-\noise)S = \sum_{C \in \C} p_C C.
\end{equation}
We will prove this latter statement.
Define the {\em Clifford polytope}
\begin{equation}\label{polytope}
P:={\rm conv}({\C})=\left\{S\mid S=\sum_{C\in {\C}}p_C C,p_C\ge 0,\ \sum_{C\in {\C}}p_C=1\right\}
\end{equation}
as the convex hull of the 24 Clifford rotation  matrices in $\rotspace$.
We have to prove
\begin{equation}\label{toprove}
(1-\noise) S \in P \ \mbox{ for any } S \in SO(3).
\end{equation}
For this we use the fact that the Clifford polytope can be alternatively described
by its facet description:
\begin{equation}
\label{eqCs}
P=\left\{S\in \rotspace\mid  \ip{F}{S}\le 1 \ \mbox{ for all } F\in {\cal F}\right\},
\end{equation} where
\begin{eqnarray}
\label{eqFB}{\cal F}:= \left\{C_1BC_2| C_1, C_2 \in \C, B\in \{B_1,B_1^T,B_2\}\right\},\\
\nonumber B_{1}:=\left(\begin{array}{ccc} 1& 0& 0 \\ 1& 0& 0 \\1& 0& 0\end{array}\right),~
B_2:=\left(\begin{array}{ccc} 1& -1& 0 \\ 1& 1& 0 \\0& 0& -1\end{array}\right).
\end{eqnarray}
One can use the software from \cite{KF} for computing the facet description (\ref{eqCs});
 we will give a direct proof in the Appendix.
In view of (\ref{eqCs}), our
 claim (\ref{toprove}) is equivalent to
\begin{equation}\label{toprove1}
(1-\noise)\ip{F}{S} \le 1 \ \mbox{ for all }S\in SO(3),\ F \in {\cal F}.
\end{equation}
Let $F\in {\cal F}$ of the form $F=C_1BC_2$ where $C_1,C_2\in {\C}$. As
$\ip{F}{S} =\ip{C_1^TSC_2^T}{B} $ and $C_1^TSC_2^T\in SO(3)$,
(\ref{toprove1}) is equivalent to
\begin{equation}\label{toprove2}
\ip{S}{B}\le {1\over 1-\noise}=2\sqrt 2 -1  \ \mbox{ for all } B\in\{B_1,B_2\},\ S\in SO(3).
\end{equation}
The case $B=B_1$ is easy to handle: For $S\in SO(3)$,
$\ip{S}{B_1}=\sum_{i=1}^3S_{i1}\le \sqrt 3 < 2\sqrt2 -1$.
We now show (\ref{toprove2}) for $B=B_2$.
Write $S \in \rotspace$ as
\begin{equation}
\label{eqU}
 S=\left(\begin{array}{ccc} a_1& a_2& a_3 \\ b_1& b_2& b_3 \\c_1& c_2& c_3\end{array}\right).
\end{equation}
Well-known necessary and sufficient conditions for $S \in SO(3)$ are
\begin{eqnarray}
\label{eqCond}
\mathbf a^T\mathbf b=0, \ \mathbf c= \mathbf a \times \mathbf b, \
\mathbf a^T\mathbf a=1, \ \mathbf b^T\mathbf b=1,
\end{eqnarray}
where $\times$ denotes the vector product, defined as
$$\mathbf a \times \mathbf b:=(a_2b_3-a_3b_2,a_3b_1-a_1b_3,a_1b_2-a_2b_1)^T.$$
Recall that, for $\mathbf a, \mathbf b, \mathbf c$ as in (\ref{eqCond}),
$\mathbf a = \mathbf b\times \mathbf c \mbox{ and }  \mathbf b =\mathbf c \times \mathbf a.$
Using $c_3=a_1b_2-a_2b_1$, we obtain $\ip{B}{S}=a_1 -a_2+ b_1 +b_2 -a_1b_2+a_2b_1$.
Therefore our task is now to prove that the optimum value of the program
\begin{equation}\label{toprove3}
\begin{array}{ll}
\max & f:=a_1 -a_2+ b_1 +b_2 -a_1b_2+a_2b_1 \\
\mbox{s.t. } &
g_1:=a_1^2+a_2^2 +a_3^2= 1\\
&    g_2:=b_1^2+b_2^2+b_3^2 = 1\\
&    g_3:=a_1b_1+a_2b_2+a_3b_3=0
    \end{array}
    \end{equation}
    is at most $2\sqrt2-1$; we in fact show that $\max f=2\sqrt2-1$.
For this, consider a global maximizer $(a,b)$ to the program (\ref{toprove3}).
Then, the Karush-Kuhn-Tucker conditions have to be satisfied,
since the gradient vectors $\{\nabla g_i(a,b)\mid i=1,2,3\}$ are linearly independent; see, e.g.,
Theorem 12.1 in \cite{NocedalWright}. (Here, the gradient vector $\nabla g_i(a,b)$
consists
of the partial derivatives with respect to the six variables
$a_1,\ldots,b_3$.)
That is, there exist scalars $\lambda_1,\lambda_2,\lambda_3$ for which
$$\nabla f (a,b) +\sum_{i=1,2,3}\lambda_i \nabla g_i(a,b)=0. $$
Equivalently, considering the partial derivatives first with respect to $(a_1,a_2,a_3)$ and then
with respect to $(b_1,b_2,b_3)$
$$\begin{array}{ccccccccc}
    \left(\begin{array}{c}1-b_2\\-1+b_1\\0 \end{array}\right)   &+& 2\lambda_1 \mathbf a   & & &+&\lambda_3 \mathbf b&=&0\\
    \left(\begin{array}{c}1+a_2\\1-a_1\\0 \end{array}\right)    & &             &+&2\lambda_2\mathbf b   &+&\lambda_3 \mathbf a&=&0.
          \end{array}
$$
Multiplying the first and the second line by $\mathbf c^T = (\mathbf a\times \mathbf b)^T$ (recall that $\mathbf c \perp \mathbf a,\mathbf b$) we get
\begin{eqnarray*}
0=c_1(1-b_2)+c_2(-1+b_1)=c_1-c_2+a_3\\
0=c_1(1+a_2)+c_2(1-a_1)=c_1+c_2+b_3.
\end{eqnarray*}
Adding (resp. subtracting) these equations yields $2c_1=-a_3-b_3$ and $2c_2=a_3-b_3$. Squaring
these two equations and then adding them gives
$2a_3^2+2b_3^2=4c_1^2+4c_2^2$. Since the rows and columns in $S$ are normalized,
we get $2(1-c_3^2)=4(1-c_3^2)$, from which we conclude $c_3^2=1$ and, therefore, $a_3=b_3=c_1=c_2=0$.
This implies $a_1^2+b_1^2=1=a_1^2+a_2^2$ and thus   $|b_1|=|a_2|$.
Similarly one can establish $|a_1|=|b_2|$.
On the basis of this observation we distinguish three cases.

\begin{enumerate}
\item $a_1=b_2=0$. Then, $|a_2|=|b_1|=1$ and $f=-a_2+b_1+a_2b_1\leq 1$.
\item $a_1\neq 0$ and $a_1=-b_2$. From $a^Tb=0$ we have  $a_1(b_1-a_2)=0$, which gives $a_2=b_1$. Then,
$f = a_1-a_2+a_2-a_1 +a_1^2+a_2^2=1$.
\item $a_1\neq 0$ and $a_1=b_2$. From $a^Tb=0$ we have  $a_1(b_1+a_2)=0$, which gives $a_2=-b_1$. Then, $f= a_1-a_2-a_2+a_1-a_1^2-a_2^2=2(a_1-a_2)-1$,
which (under the condition $a_1^2+a_2^2=1$) is clearly maximized by $a_1=-a_2=1/\sqrt{2}$.
Therefore, we find $\max f = 2\sqrt{2} -1$.
\end{enumerate}
Thus,  we have shown that the optimum value of the program (\ref{toprove3}) is equal to
$2\sqrt{2} -1$, which concludes the proof.
\end{proof}

\begin{mylemma}
\label{lemmaqMain}
Let $f : \{0,1\}^{m_A} \times \{0,1\}^{m_B} \rightarrow \{0,1\}$ be a function and $K$ a quantum circuit for $f$ with correctness probability $c> 1/2$ which uses only gates from \Cln and measurements in the computational basis. Then the \emph{randomized} communication complexity of $f$ is at most one bit.
\end{mylemma}
\begin{proof}
From Lemma \ref{lemmaMain} we know how two  parties, Alice and Bob, can simulate perfect Clifford gates. From Lemma \ref{lemmaSimulate} we know how they can jointly simulate the other noisy 1-qubit gates in $\Cln$, where they use shared randomness to make sure that they always simulate the same Clifford gate.
\end{proof}


We can now prove an upper bound on the noise in fault-tolerant quantum computation.
\begin{mytheorem}
\label{theMain}
The set of gates from \Cl\ together with 1-qubit gates with depolarizing noise more than \noise\  and one single-qubit measurement is not sufficient for arbitrary classical computation.
\end{mytheorem}
\begin{proof}
The result follows by Lemma \ref{lemmaqMain} and the fact that there are functions with communication complexity greater than $1$.
In fact we have that none of the functions with unbounded error communication complexity $C_f^{unbounded}>1$ can be computed.
\end{proof}

\section{Discussion and extensions}
\paragraph{Best gates}
\label{remPI8}
 From the proof of Lemma \ref{lemmaSimulate} we see that the rotation matrix $S$ which achieves the optimal value, is
\begin{equation}
 \left(\begin{array}{ccc} 1/\sqrt{2}&-1/\sqrt{2}&0 \\ 1/\sqrt{2}& 1/\sqrt{2}& 0 \\0& 0& -1\end{array}\right).
\end{equation}
Multiplying from the right by the Clifford-matrix $diag(1,-1,-1)$ we get a rotation around the $z$-axis by $\pi/4$.   The $\pi/8$-gate
$$T=\left(\begin{array}{cc} \exp(-i\pi/8) &0 \\0& \exp(i\pi/8) \end{array}\right)$$ performs a rotation of  $\pi/4$ around the $z$-axis. So, the $\pi/8$-gate and its symmetric versions
are the ones which need the most depolarizing noise to be simulated by gates from \Cl.

\paragraph{Worst case noise}
In Lemma \ref{lemmaSimulate}  we asked with how much depolarizing noise all 1-qubit unitary gates are equivalent to probabilistic mixtures of Clifford gates. Similarly to \cite{VirmaniHuelgaPlenio04} one can also ask how much arbitrary noise is needed to make every gate a mixture of Cliffords. More precisely what is the value $\tilde \theta=\sup_{U \in SU(2)} p_U$ , where $p_U$ is the infimum of all $p$ such that there is a completely positive trace-preserving 1-qubit quantum operation ${\cal E}_U$ with the property that the noisy implementation of $U$
$$U': \rho \mapsto (1-p) U\rho U^\ast + p{\cal E}_U(\rho) $$
becomes a probabilistic mixture of Clifford operation.

In this section we will provide some bounds on $\tilde \theta$. Let $K \in SU(2)$ be any operation that in Bloch representation maps the state $X$-eigenstate $v_X=(1,0,0)$ to $u=\frac{1}{\sqrt{3}}(1,1,1)$. Note that a probabilistic mixture of 1-qubit Clifford operations $C=\sum_ip_i C_i$ can map $v_X$ only into the octahedron $\cal O$ spanned by $v_X=(1,0,0)$, $v_Y=(0,1,0)$ and $v_Z=(0,0,1)$ and their negatives $-v_X,-v_Y,-v_Z$ (see also \cite{BK:Magic}). 
Note that the state of $\cal O$  which is closest to $u$ is $\frac{1}{3}(1,1,1)=\frac{1}{\sqrt{3}}u$ and their distance is  $||u-1/\sqrt{3}u||_2=1-\frac{1}{\sqrt{3}}$. The Bloch-state which is furthest away from $u$ is $-u$. All three of these states lie on a line. With this it is clear that the state $u_{noise}$ which needs the smallest noise $p$, such that $(1-p)u +pu_{noise}$ 
is inside the octahedron is $-u$ and the optimal $p$ is $\frac{1}{2}(1-\frac{1}{\sqrt{3}})$. This implies $21\% \approx \frac{1}{2}(1-\frac{1}{\sqrt{3}})\le \tilde\theta$. 

To get an upper bound, recall that by Lemma \ref{lemmaSimulate} for any gate $U \in SU(2)$ the operation
$$U': \rho \mapsto (1-p) U \rho U^\ast + p \id/2$$
is a Clifford  operation, if $p \ge \noise$.  Setting ${\cal E}_U(\rho) =\frac{1}{3}\left(X U \rho U^\ast X + Y U\rho U^\ast Y + Z U\rho U^\ast Z\right) $ and noting that for any 1-qubit density matrix it holds
$\id/2= \frac{1}{4}\left(\rho + X \rho X + Y \rho Y + Z \rho Z\right)$ we can rewrite the action of $U'$ also as
$$U': \rho \mapsto (1-\frac{3}{4}p)U \rho U^\ast + \frac{3}{4}p{\cal E}_U(\rho).$$
Thus $\tilde \theta \le  \frac{3}{4} \noise \approx 34 \%$. Note that this is certainly not tight, since all gates, apart from the $\pi/8$-gate (and its symmetric versions), need less than $\noise~$ depolarizing noise to make it a probabilistic mix of Clifford operations, which implies they need less than $\frac{3}{4}\noise~$ worst case noise. However, as follows from \cite{VirmaniHuelgaPlenio04}, the worst case noise for the $\pi/8$-gate(s) is only $\frac{1}{2}-\frac{1}{2\sqrt{2}}\approx 15\%$.

We leave it as an interesting open question to determine the precise value of $\noise$.

\paragraph{Classical Co-processing}
Theorem~\ref{theMain} states that fault tolerant quantum computing
is not possible if we have depolarizing noise at least  $\noise \approx 45 \%$ on one qubit gates even if we can use perfect
gates from $\Cl$ in our fault tolerant circuit design. Is this
optimal? Could it be that with less than $\noise$ noise on the
single qubit gates and perfect gates from $\Cl$ still
no fault tolerant circuit design is possible. We leave this as an open
question, but Ben Reichardt~\cite{R:personal05} pointed out that 
when we allow perfect classical computation in addition to  perfect gates from
$\Cl$ and perfect measurements in the computational basis, for any quantum circuit one can
build a fault tolerant quantum circuit, that tolerates noise less than
$\noise$ on single qubit gates. This fault tolerent
implementation has only a constant factor slowdown in time.


The argument builds upon magic-state distillation, introduced in
\cite{BK:Magic}, and goes as follows. Assume we have at our
disposal   noisy $\pi/8$-gates $T'$, with depolarizing noise strictly less
than $\tilde \theta$,  i.e.  $T'(\rho)=(1-p)T\rho
T^\ast+p\id/2$ with $p<\noise$, where  $T$ is the perfect
$\pi/8$ gate. Then apply $T'$ to the second half
of an EPR-pair and measure the observable $Z \otimes Z$, which can be
implemented as a measurement in the computation basis with additional
gates from $\Cl$. If the
outcome is  $-1$ throw away the state and do the experiment again.
If the outcome is $+1$, apply a CNOT from the first to the second
qubit, which gives
\begin{equation}\label{magic}
\frac{1}{2}\left(\id+
    \frac{1-p}{1-p/2}\frac{1}{\sqrt{2}}X +
    \frac{1-p}{1-p/2}\frac{1}{\sqrt{2}}Y   \right)  \otimes
  |0\rangle\langle0|.
\end{equation}
Using the result from \cite{R:Magic} an arbitrary supply of qubits in the state of the first qubit of~(\ref{magic}) can be used to distill magic states in the $H$-direction, which together with stabilizer operations is sufficient for quantum computation. We do not know if this also holds for other than the $\pi/8$-gate.

Note the this is tight for the $\pi/8$-gate, since stabilizer operations (Cliffords, measurements in the computational basis and classical co-processing)together with $\pi/8$-gates with depolarizing noise \noise\ can be efficiently simulated 
classically, as follows from our Lemma \ref{lemmaSimulate} and the Gottesman-Knill Theorem.



\paragraph{Allowing some perfect unitaries}
Our threshold theorem says the following. Let $f$ be a function such
that  it requires more than one bit of communication in order to
compute it, when the input bits are partitioned over Alice and
Bob. There is no quantum circuit consisting of {\em perfect} Clifford
operations and single qubit gates with noise \noise~  ($\approx 45\%$) that can
compute $f$. We can strengthen this result to allow a small amount
of perfect single qubits as well. Assume that $f$ requires $m$ bits of
communication to be computed. There is no quantum circuit that uses
perfect Clifford operations, $s$ perfect single qubit gates, and
single qubit gates with noise \noise~ that computes $f$, for $2s +1  < m$.
The reason we get this strengthening is because in our simulation,
Lemmas 1 and 2, Alice sends to Bob  whenever he wants to perform a
perfect single qubit gate on some qubit, her classical share  $a$ and
$b$ of that specific qubit. Bob can now perform the perfect qubit gate
on that qubit and
they proceed as in Lemma 1 and 2. By the end of the simulation Alice has
sent $2s +1$ bits to Bob and he will be able to compute $f$,
contradicting that the communication complexity of $f$ is at least
$m> 2s +1$.


\section{Acknowledgements}
We thank Neboj{\v s}a Gvozdenovi\'c and Hartwig Bosse and also Troy Lee, Ben Reichardt and  Stephanie Wehner for useful 
discussions and Komei Fukuda for providing the software {\tt cdd+} \cite{KF}.
We also thank Scott Aaronson and Daniel Gottesman for discussions about their work.
We are also grateful to the Newton Institute, Cambridge, where this work was started.

\bibliographystyle{plain}


\newpage
\appendix
\section{Computing the facets of the Clifford polytope}
\label{secFacets}
We give here the facet description for the Clifford polytope $P$ defined in (\ref{polytope})
as the convex hull of the set $\C$ consisting of the 24 Clifford rotations of the 3-space.
Define the polyhedron
$$Q:=\{S\in \rotspace\mid \ip{F}{S}\le 1\ \mbox{ for all } F\in {\cal F}\},$$
where $\cal F$ is as in (\ref{eqFB}). Our objective is to show the equality $P=Q$.

To start with, let us prove the easy inclusion $P\subseteq Q$. For this, let $C\in {\C}$ and
$F\in {\cal F}$ of the form $F=C_1BC_2$ with $C_1,C_2\in {\C}$ and $B\in \{B_1,B_1^T,B_2\}$.
Then, $\ip{F}{C}=\ip{B}{C_1^TCC_2^T}$. As $C_1^TCC_2^T\in {\C}$,
it suffices to verify that $\ip{B}{C}\le 1$ for any $C\in {\C}$ and $B=B_1,B_2$. (We have used here the fact that
$\C$ is a group which is closed under transposing matrices.)
For $C\in {\C}$, the inequality $\ip{B_1}{C}\le 1$ is obvious and the inequality
$\ip{B_2}{C}\le 1$ can be checked by direct inspection.

The reverse inclusion $Q\subseteq P$ follows
from the following result.

\begin{mytheorem}\label{theo0}
Any facet of the polytope $P$ is defined by an
inequality
of the form $\ip{F}{S}\le 1$ where  $F\in {\cal F}$.
\end{mytheorem}

The rest of the Appendix is devoted to the proof of this result.
We first need to go in more detail into the structure of the Clifford matrices.

\subsection{Preliminaries about the Clifford matrices}
\label{subsecRotMat}
Each matrix $C\in \C$ corresponds to a ``signed permutation" $(\sigma,s)$, where
$\sigma\in \Sym(3)$ and $s\in \{\pm 1\}^3$. Namely, $C$ has nonzero entries precisely at the $(\sigma(i),i)$-positions
with $C_{\sigma(i),i} =s_i$ for $i=1,2,3$; we then also denote $C$ as $C_{\sigma,s}$.
The condition $\det(C)=1$ translates into $s_1s_2s_3= \sign(\sigma)$; that is,
$s_1s_2s_3= 1$ if $\sigma$ is an  even  permutation (i.e.,
one of $\sigma_1:=(1,2,3)$, $\sigma_2:=(2,3,1)$, $\sigma_3:=(3,1,2)$) and
$s_1s_2s_3= -1$ if $\sigma$ is an odd permutation  (i.e.,
one of $\sigma_4:=(1,3,2)$, $\sigma_5:=(2,1,3)$, $\sigma_6:=(3,2,1)$).
Thus the set $\C$ of Clifford matrices is naturally partitioned into six subclasses
$$\C=\displaystyle\bigcup_{\sigma\in \Sym(3)} \C_\sigma, \mbox{ where }
\C_\sigma:=\{C_{\sigma,s}\mid s\in \{\pm 1\}^3,\ s_1s_2s_3=\sign(\sigma)\}$$
with $|\C_\sigma|=4$.
For convenience we display in the table below the six subclasses $\C_\sigma$;
the nonzero entries are indicated by  $*$.
$$\begin{array}{c|c}
\mbox{Even permutations} & \mbox{Odd permutations}\\\hline

\sigma_1=(1,2,3) \ \ \ \C_{\sigma_1}:
\left(
\begin{array}{ccc}
* & 0 & 0 \\0& *& 0\\ 0& 0& *
\end{array}
\right)&
\sigma_4=(1,3,2) \ \ \ \C_{\sigma_4}:\left(
\begin{array}{ccc}
* & 0 & 0 \\ 0 & 0& *\\ 0& *& 0
\end{array}
\right)  \\

\sigma_2=(2,3,1) \ \ \ \C_{\sigma_2}: \left(
\begin{array}{ccc}
0 & 0 & * \\ * & 0& 0\\ 0& *& 0
\end{array}
\right)& \sigma_5=(2,1,3) \ \ \ \C_{\sigma_5}:\left(
\begin{array}{ccc}
0 & * & 0 \\ * & 0& 0\\ 0& 0& *
\end{array}
\right)
 \\

\sigma_3=(3,1,2) \ \ \ \C_{\sigma_3}:\left(
\begin{array}{ccc}
0 & * & 0 \\0& 0& *\\ *& 0& 0
\end{array}
\right) &
\sigma_6=(3,2,1) \ \ \ \C_{\sigma_6}:\left(
\begin{array}{ccc}
0 & 0 & * \\0& *& 0\\ *& 0& 0
\end{array}
\right) \\

\end{array}
$$
\begin{center} Table 1 \end{center}

The following observation can be directly verified and will be useful for the proof.

\begin{myob}\label{ob0}
Let $\sigma\in \Sym(3)$. Then, $\sum_{C\in \C_\sigma}C=0$. Moreover,
for any position
$(\sigma(i),i)$ corresponding to a nonzero entry for matrices in $\C_\sigma$ and for $d\in\{\pm 1\}$, there exist
$C,C'\in \C_\sigma$ with $C+C' = 2dE_{\sigma(i),i}$, which implies
$d E_{\sigma(i),i}\in P$. Thus, $\pm E_{i,j}\in P$ for any $i,j=1,2,3$.
\end{myob}

We now proceed with the proof of Theorem \ref{theo0}.
Let $\ip{F}{S}\le b$ be an inequality defining a facet of $P$, where
$F\in \rotspace$ and $b\in \rn$. That is,
 the inequality $\ip{F}{S}\le b$ is valid for $P$, which means that
$\ip{F}{S}\le b$ holds for any $S\in P$, and  the set
$${\cal R}_F:=\{C\in {\C}\mid \ip{F}{C}=b\}$$
contains nine affinely independent matrices.
Without loss of generality, we may assume that $b=1$. Indeed,
$b\ge 0$ since $0\in P$. Moreover, $b>0$ for, otherwise, we would have $F_{ij}=0$ for all $i,j=1,2,3$, implying  $F=0$, in view
of Observation \ref{ob0}.  Thus, by rescaling, we can now
assume that the facet is of the form $\ip{F}{S}\le 1$.
We sometimes speak of the ``facet $F$" for short.
Our objective is to show
that $F=C_1BC_2$ for some $C_1,C_2\in {\C}$, $B\in\{B_1,B_1^T,B_2\}$.

Call $F,F'\in \rotspace$ {\em equivalent} if $F'=C_1FC_2$ for some $C_1,C_2\in \C$.
Then, as $\C$ is a group, $\ip{F'}{S}\le 1$ defines a facet
of $P$ if and only if $\ip{F}{C}\le 1$ does.
Moreover, $\R_{F'}=C_1\R_FC_2=\{C_1CC_2\mid C\in \R_F\}$.
This property will be used repeatedly throughout  the proof  as it permits  to
exploit symmetry and
to reduce the  number of case checking.

The proof is based on a detailed inspection of the structure of the set
$\R_F$.
We begin with collecting several properties of the matrix $F$  and the set $\R_F$.

\begin{myob} \label{ob1}
$|\R_F \cap \C_\sigma|\le 3$ for any $\sigma\in \Sym(3)$.
\end{myob}
\begin{proof}
If $\C_\sigma\subseteq \R_F$, then $\ip{F}{C}=1$ for any $C\in \C_\sigma$, which implies
$4=\sum_{C\in \C_\sigma}\ip{F}{C}$, contradicting the fact that $\sum_{C\in \C_\sigma}C=0$ by Observation \ref{ob0}.
\end{proof}


\begin{myob}
\label{ob2}
If $F_{ij}=d\in \{-1,1\}$,  then all $C \in \C$ with $C_{ij}=d$ belong to  $R_F$.
\end{myob}
\begin{proof}
Let $C\in\C$ with $C_{ij}=d$. There exists $C'\in \C$ with
$C+C' = 2dE_{ij}$. Summing  $\ip{F}{C}\le  1$ and $\ip{F}{C'}\le 1$ yields
$\ip{F}{C+C'}\le 2$. As $\ip{F}{C+C'}= 2dF_{ij}=2$, we have the equalities
$\ip{F}{C}=\ip{F}{C'}=1$, which implies   $C\in \R_F$.\end{proof}

\begin{myob}
\label{ob3}
Let $C\neq C' \in \R_F \cap \C_\sigma$ (for some $\sigma\in\Sym(3)$) and assume
that $C_{\sigma(i),i}=C_{\sigma(i),i}'=d\in \{\pm 1\}$ for some $i\in \{1,2,3\}$.
Then, $F_{\sigma(i),i}=d$ and
$F_{\sigma(j),j}+\sign(\sigma)d F_{\sigma(k),k}=0$ with
$\{j,k\}=\{1,2,3\}\setminus~\{i\}$.
\end{myob}

\begin{proof}
Equality $F_{\sigma(i),i}=d$ follows from the fact that $C+C'=2dE_{\sigma(i),i}$.
Then, $1=\ip{F}{C}$ implies $0=F_{\sigma(j),j}C_{\sigma(j),j} +F_{\sigma(k),k}C_{\sigma(k),k}$.
Using $C_{\sigma(i),i}C_{\sigma(j),j}C_{\sigma(k),k}=\sign(\sigma)$, we find
$F_{\sigma(j),j}+\sign(\sigma)d F_{\sigma(k),k}=0$.
\end{proof}

Our last observation is an easy corollary of the former two observations.
\begin{myob}\label{ob4}
If $F_{\sigma(i),i}=d\in \{\pm 1\}$ (for some $\sigma\in \Sym(3)$),
then
$F_{\sigma(j),j} +\sign(\sigma)d F_{\sigma(k),k}=0$, where $\{i,j,k\}=\{1,2,3\}$.
\end{myob}

One can verify that, for $F=B_1,B_1^T$, $|\R_F\cap\C_\sigma|= 2$ for all $\sigma\in\Sym(3)$
while, for $F=B_2$,  $|\R_F\cap\C_\sigma|=3$ for  $\sigma=\sigma_1,\sigma_5$.
Based on this observation we now distinguish two cases: Either,
 $|\R_F\cap\C_\sigma|\le 2$ for all $\sigma\in\Sym(3)$ (in which case we show that
$F$ is equivalent to $B_1$ or $B_1^T$), or  $|\R_F\cap\C_\sigma|=3$ for some $\sigma \in\Sym(3)$
(in which case we show that $F$ is equivalent to $B_2$).

\subsection{The case $|\R_F\cap \C_\sigma|=3$ for some $\sigma\in\Sym(3)$}
\label{app3pts}

Using symmetry, we may assume that $|\R_F \cap \C_\sigma|=3$ for the (odd)
permutation $\sigma=\sigma_4$.
We prove this in detail to show how this kind of symmetry argument works.
Define the matrices
\begin{equation}\label{3matC}
C_1=\left(\begin{matrix} - & 0 & 0\cr 0 & 0 & - \cr 0 & - & 0\end{matrix}\right),\
C_2=\left(\begin{matrix} 0 & + & 0 \cr 0 & 0 & +\cr + & 0 & 0\end{matrix}\right),\
C_3=\left(\begin{matrix} 0 & 0 & +\cr + & 0 & 0\cr 0  & + & 0\end{matrix}\right)
\end{equation}
lying, resp., in $\C_{\sigma_4},$ $\C_{\sigma_3}$, $\C_{\sigma_2}$.
Recall that $+,-$ stand for $1,-1$, respectively.
Our assumption is that $|\R_F \cap \C_{\sigma_i}|=3$ for some $i=1,\ldots,6$; we show that one can replace $F$ by another equivalent facet $F'$ in such a way that $i=4$ holds.
For this, suppose first $i=2,3$.
As the mapping $X\mapsto X C_i$ maps
$\C_{\sigma_i}$ to $\C_{\sigma_1}$, we can replace the facet $F$ by $F':=FC_i$ and then  we find
$|\R_{F'}\cap \C_{\sigma_1}|=3$ since $\R_{F'}=\R_FC_i$. Thus we may assume $|\R_F\cap\C_{\sigma_1}|=3$.
As the mapping
$X\mapsto XC_1$ maps $\C_{\sigma_1}$ to $\C_{\sigma_4}$, replacing the facet $F$ by $F':=FC_1$,
we find $|\R_{F'}\cap \C_{\sigma_4}|=3$. Thus we can now assume
$|\R_F\cap\C_{\sigma_i}|=3$ for some $i=4,5,6$.
If $i=5$, as the mapping $X\mapsto X C_3$ maps $\C_{\sigma_5}$ to
$\C_{\sigma_4}$, replace $F$ by $F':=FC_3$; if $i=6$, the mapping $X\mapsto X C_2$ maps
$\C_{\sigma_6}$ to $\C_{\sigma_4}$ and one can replace $F$ by $F':=FC_2$; in both cases we get back to the case when
$|\R_{F'}\cap\C_{\sigma_4}|=3$.

Thus we now assume $|\R_F\cap\C_{\sigma_4}|=3$. Moreover, we may assume that
the following matrices from $\C_{\sigma_4}$
\begin{eqnarray}
\label{3matrices}
\left(
\begin{array}{ccc}
+ & 0 & 0 \\0& 0& -\\ 0& +& 0
\end{array}
\right),\
\left(
\begin{array}{ccc}
- & 0 & 0 \\0& 0& +\\ 0& +& 0
\end{array}
\right),\
\left(
\begin{array}{ccc}
- & 0 & 0 \\0& 0& -\\ 0& -& 0
\end{array}
\right)
\end{eqnarray}
belong to $\R_F$. (To see this, replace if necessary $F$ by $FC$, where $C\in\C_{\sigma_1}$.)
Using Observation \ref{ob3},  we obtain  $F_{11}=-1, F_{23}=-1, F_{32}=1$.
 From this we get by Observation \ref{ob2} that also the matrices
\begin{eqnarray}
\label{U1}
\left(
\begin{array}{ccc}
- & 0 & 0 \\0& -& 0\\ 0& 0& +
\end{array}
\right), \
\left(
\begin{array}{ccc}
- & 0 & 0 \\0& +& 0\\ 0& 0& -
\end{array}
\right)\in\C_{\sigma_1},  \\
\label{U2}
\left(
\begin{array}{ccc}
0 & 0 & + \\+& 0& 0\\ 0& +& 0
\end{array}
\right),\
\left(
\begin{array}{ccc}
0 & 0 & - \\-& 0& 0\\ 0& +& 0
\end{array}
\right)\in\C_{\sigma_2},\\
\label{U3}
\left(
\begin{array}{ccc}
0 & + & 0 \\0& 0& -\\ -& 0& 0
\end{array}
\right),\
\left(
\begin{array}{ccc}
0 & - & 0 \\0& 0& -\\ +& 0& 0
\end{array}
\right)\in\C_{\sigma_3}  \\
\end{eqnarray}
belong to  $\R_F$.
By Observation \ref{ob3},  we also obtain $F_{22}=F_{33}$, $F_{12}=F_{31}$ and $F_{13}=-F_{21}$.

\begin{myclaim}
There exists also an even permutation $\sigma\in\Sym(3)$ for which
$|\R_F\cap\C_\sigma|=3$.
\end{myclaim}

\begin{proof}
Assume for  contradiction that, for $i=1,2,3$,
 the set $\R_F \cap \C_{\sigma_i}$ contains only the respective two matrices from (\ref{U1})-(\ref{U3}).
 Choose a  subset $\B\subseteq \R_F$ consisting of nine affinely independent matrices and such that
$\R_F\cap\C_{\sigma_4}\subseteq \B$.
We have $|\B\cap\C_{\sigma_1}|\le 1$, since the two matrices in (\ref{U1}) are affinely dependent with the last two matrices
in (\ref{3matrices}).  Similarly, $|\B\cap\C_{\sigma_2}|\le 1$,
 $|\B\cap\C_{\sigma_3}|\le 1$. As $|\B|=9$, we deduce that
$|\B\cap\C_{\sigma_5}|\ge 2$ or $|\B\cap\C_{\sigma_6}|\ge 2$.
Assume first that $|\B\cap \C_{\sigma_5}|\ge 2$.
Say, $C\ne C'\in \R_F\cap\C_{\sigma_5}$.
Then $C$ and $C'$ have the same nonzero entry $d \in \{-1,1\}$ in some position $(k,l)$.
By Observation \ref{ob3} this yields  $F_{kl}=d$. Now, there is also an even
permutation $\sigma$ for which $k=\sigma(l)$.
By Observation \ref{ob2} we then deduce that at least two matrices from $\C_\sigma$
must be in $\R_F$, which contradicts our assumption.
The other case $|\B\cap \C_{\sigma_6}|\geq 2$ goes analogously.
\end{proof}

It is sufficient to consider the case $|\R_F\cap\C_{\sigma_1}|=3$. Indeed,
if $|\R_F\cap\C_{\sigma_2}|=3$, then
one may replace $F$ by $C_3FC_4$ with $C_3$ as in (\ref{3matC}) and
$$C_4:=\left(\begin{matrix} 0 & 0 & + \cr - & 0 & 0 \cr 0 & - & 0\end{matrix}\right),$$
since
 the mapping $X\mapsto C_3XC_4$ maps $\C_{\sigma_2}$ to $\C_{\sigma_1}$ and
 preserves the set of three matrices from
(\ref{3matrices}), as well as the set of 6 matrices from (\ref{U1})-(\ref{U3}) (namely,
(\ref{U1}) $\rightarrow $ (\ref{U3}) $\rightarrow $ (\ref{U2}) $\rightarrow $ (\ref{U1})).
 One can handle the case when
$|\R_F\cap \C_{\sigma_3}|=3$ in the same way.

The set $\R_F$ already contains the matrices
$$D_1:=\left(\begin{matrix} - & 0 & 0 \cr 0 & - & 0 \cr 0 & 0 & +\end{matrix}\right),\
D_2:=\left(\begin{matrix} - & 0 & 0 \cr 0 & + & 0 \cr 0 & 0 & -\end{matrix}\right)$$
from $\C_{\sigma_1}$ (displayed in  (\ref{U1})). The remaining two matrices of
$\C_{\sigma_1}$ are
$$D_3:=\left(\begin{matrix} + & 0 & 0 \cr 0 & + & 0 \cr 0 & 0 & +\end{matrix}\right),\
D_4:=\left(\begin{matrix} + & 0 & 0 \cr 0 & - & 0 \cr 0 & 0 & -\end{matrix}\right).$$
If $D_4\in \R_F$, one may replace the facet $F$ by $F':=D_2FD_1$ to obtain that
$D_1,D_2,D_3\in \R_{F'}$, since the mapping
$X\mapsto D_2XD_1$ maps $\{D_1,D_2,D_4\}$ to $\{D_1,D_2,D_3\}$ and leaves the set of 3 matrices from
(\ref{3matrices}) invariant as well as the set of 6 matrices from (\ref{U1})-(\ref{U3}).
Thus we may assume that $D_3\in \R_F$.

By Observation \ref{ob3}, we find that $F_{33}=F_{22}=1$.
As $F_{22}=1$, Observation \ref{ob4} implies that  $F_{31}=F_{13}$.
Similarly, $F_{33}=1$ implies that $F_{12}=F_{21}$.
Putting all equations together we obtain  $F_{12}=F_{21}=-F_{13}=-F_{31}=-F_{12}$, implying they
 are all zero. Thus
\begin{equation}
F=\left(
\begin{array}{ccc}
- & 0 & 0 \\0& + & -\\ 0& +& +
\end{array}
\right)=\left(\begin{matrix} 0 & 0 & 1\cr 1 & 0 & 0\cr 0 & 1 & 0\end{matrix}\right)
B_2 \left(\begin{matrix} 0 & 1 & 0\cr 0 & 0 & 1\cr 1 & 0 & 0\end{matrix}\right).
\end{equation}

\subsection{The case $|\R_F \cap \C_\sigma|\le 2$ for all $\sigma\in\Sym(3)$}
\label{app2pts}

Let again $\B\subseteq \R_F$ consist of nine affinely independent matrices.
As $|\R_F|\ge 9$, $|\R_F\cap\C_\sigma|=2$ for at least three permutations $\sigma$.
W.l.o.g. we can assume that two of those permutations are odd permutations and that they are
equal, say, to $\sigma_4$ and $\sigma_6$ (replacing if necessary $F$ by an equivalent facet).
Further we may assume $\R_F$ contains the following two matrices of $\C_{\sigma_4}$:
\begin{equation}\label{BU1}
\left(\begin{matrix} - & 0 & 0\cr 0 & 0 & -\cr 0 & - & 0\end{matrix}\right),\
\left(\begin{matrix} - & 0 & 0\cr 0 & 0 & +\cr 0 & + & 0\end{matrix}\right)\in \R_F.\end{equation}
This can be seen using the following two mappings
$X\mapsto C_2XC_2$ (with $C_2$  defined as in (\ref{3matC})) and $X\mapsto CX$
(with $C\in \C_{\sigma_1}$) which permit to map any subset of size 2 of $\C_{\sigma_4}$
to any other such subset and which preserve $\C_{\sigma_6}$ as well.
We choose the basis $\B$ containing the two matrices of (\ref{BU1}).
>From Observation \ref{ob3} we find $F_{11}=-1$ and $F_{23}=-F_{32}\ne \pm 1$;
the latter inequality follows from the fact that $|\R_F\cap\C_{\sigma_4}|=2$ combined with
Observation \ref{ob2}.
As $F_{11}=-1$, by Observation \ref{ob2},
\begin{equation}
\label{BU4}
\left(
\begin{array}{ccc}
- & 0 & 0 \\0& - & 0\\ 0& 0& +
\end{array}
\right), \
\left(
\begin{array}{ccc}
- & 0 & 0 \\0& + & 0\\ 0&0& -
\end{array}
\right)\in \R_F\cap\C_{\sigma_1}
\end{equation}
and Observation \ref{ob3} implies $F_{22}=F_{33}\ne\pm 1$.
At most one of the two matrices in (\ref{BU4}) belongs to $\B$ since they are affinely
dependent with the matrices in (\ref{BU1}). Say, $|\B\cap\C_{\sigma_1}|=1$.

Let us now examine which two matrices of $\C_{\sigma_6}$ belong to
$\R_F$.
Set
$$C_5:=\left(\begin{matrix} + & 0 & 0\cr 0 & - & 0\cr 0 & 0 & -\end{matrix}\right)\in\C_{\sigma_1},\
X_1:=\left(\begin{matrix} 0 & 0 & - \cr 0 & - & 0\cr - & 0 & 0\end{matrix}\right)\in\C_{\sigma_6}.$$
The two mappings $X\mapsto XC_5$ and $X\mapsto C_5X$ preserve the set of matrices in
(\ref{BU1}) and permit to map any other matrix of $\C_{\sigma_6}$
to the matrix $X_1$.
Therefore we can assume w.l.o.g. that $X_1\in \R_F\cap\C_{\sigma_6}$.
The second matrix of $\R_F\cap\C_{\sigma_6}$ does not have entry $-1$ at the position
$(2,2)$ since, otherwise, $F_{22}=-1$ contradicting an earlier claim.
Hence the second matrix in  $\R_F\cap\C_{\sigma_6}$ is
$$X_2:=\left(\begin{matrix} 0 & 0 & + \cr 0 & + & 0\cr - & 0 & 0\end{matrix}\right),\
\mbox{ or }
X_3:=\left(\begin{matrix} 0 & 0 & - \cr 0 & + & 0\cr + & 0 & 0\end{matrix}\right).$$
\begin{enumerate}
\item  Consider first the case when $X_2\in  \R_F\cap\C_{\sigma_6}$.
Then, $F_{31}=-1$ and $F_{22}=-F_{13}\ne \pm 1$. As $F_{31}=-1$, we have
\begin{equation}\label{BU5}
\left(
\begin{array}{ccc}
0 & + & 0 \\0& 0 & -\\ -& 0& 0
\end{array}
\right),\
\left(
\begin{array}{ccc}
0 & - & 0 \\0& 0 & +\\ -& 0& 0
\end{array}
\right)\in \R_F\cap\C_{\sigma_3}
\end{equation}
and $F_{12}=F_{23}\ne \pm 1$.
As $\B$ contains at most three of the matrices $X_1,X_2$ and in (\ref{BU5}), we must have
$|\B\cap\C_{\sigma_2}|=2$ or $|\B\cap\C_{\sigma_5}|=2$.
We obtained eralier that
  $F_{33}=F_{22}=-F_{13}\ne \pm 1$ and $F_{12}=F_{23}=-F_{32}\ne \pm 1$.
In other words, the second and third columns of $F$ contain no entry $\pm 1$.
On the other hand, the two matrices from $B \cap \C_{\sigma_i}$ ($i=2,5$)
 have one common nonzero entry which therefore is located in the first column, at the position
$(2,1)$. This implies $F_{21}=\pm 1$.
\begin{enumerate}\item
If $F_{21}=1$, then   Observation \ref{ob4} implies
$F_{12}=F_{33}$ and $F_{13}=-F_{32}$. Combining with the former relations on entries of $F$, we find
\begin{eqnarray}
F=\left(
\begin{array}{ccc}
- & 0 & 0 \\+& 0 & 0\\ -& 0& 0
\end{array}
\right).
\end{eqnarray}
\item If $F_{21}=-1$, then in the same way  we find
\begin{eqnarray} F=\left(
\begin{array}{ccc}
- & 0 & 0 \\-& 0 & 0\\ -& 0& 0
\end{array}
\right).
\end{eqnarray}
\end{enumerate}
In both cases we find that $F$ is equivalent to $B_1$.
\item Consider now the case when $X_3\in \R_F\cap\C_{\sigma_6}$.
Then, $F_{13}=-1$, $F_{22}=-F_{31}\ne\pm 1$,
\begin{equation}\label{BU6}
\left(
\begin{array}{ccc}
0 & 0 & - \\-& 0 & 0\\ 0& +& 0
\end{array}
\right),\
\left(
\begin{array}{ccc}
0 & 0 & - \\+& 0 & 0\\ 0& -& 0
\end{array}
\right)\in \R_F\cap\C_{\sigma_2}
\end{equation}
and $F_{21}=F_{32}\ne \pm 1$.
In the same way as in the first case one finds that $F$ is equivalent to
$B_1^T$.
\end{enumerate}
This concludes the proof of Theorem \ref{theo0}.
\end{document}